\documentclass[a4paper,11pt]{article}

\pdfoutput=1 % if your are submitting a pdflatex (i.e. if you have images in pdf, png or jpg format)

\usepackage{bbm}
\usepackage{jheppub} % for details on the use of the package, please see the JHEP-author-manual
\usepackage[T1]{fontenc}
\usepackage{simplewick} % Wick contractions
\usepackage{makecell}	% Thicker table row lines. Summon with \Xhline{3\arrayrulewidth} .
\usepackage{tabu}		% Thicker table columns lines. Use tabu instead of tabular.
\usepackage[makeroom]{cancel} % gamma-tech-stuff
\usepackage[normalem]{ulem} % underline
\usepackage[utf8]{inputenc}
\usepackage{amsmath,amsfonts,amsthm,amssymb} % Math packages & curly letters
\usepackage{subcaption} % For multiple figures in one figure

\newcommand{\p}{\partial}
\newcommand{\eq}{&\quad}
\newcommand{\tq}{&\quad\quad}
\newcommand{\qRq}{\quad\Rightarrow\quad}

\newcommand{\para}{\parallel}

\newcommand{\diag}{\text{diag}}

\newcommand{\mco}{\mathcal{O}}

\newcommand{\M}{\mathcal{M}}
\newcommand{\Q}{\mathbb{Q}}

\newcommand{\R}{\mathbb{R}}
\newcommand{\Z}{\mathbb{Z}}
\newcommand{\1}{\mathbbm{1}}
\newcommand{\vph}{\varphi}
\newcommand{\vt}{\vartheta}

\newcommand{\hD}{{\hat{\Delta}}}

\newcommand{\hga}{{\hat{\gamma}}}

\newcommand{\hO}{{\hat{\mathcal{O}}}}

\newcommand{\al}{\alpha}

\newcommand{\de}{\delta}
\newcommand{\e}{\epsilon}
\newcommand{\ph}{\phi}
\newcommand{\g}{\gamma}

\newcommand{\la}{\lambda}
\newcommand{\m}{\mu}

\newcommand{\thet}{\theta}
\newcommand{\up}{\upsilon}
\newcommand{\x}{\xi}

\newcommand{\De}{\Delta}
\newcommand{\Ph}{\Phi}
\newcommand{\G}{\Gamma}

\preprint{UUITP-08/23}

\title{\boldmath The $O(N)$-flavoured replica twist defect}

\author{Alexander Söderberg Rousu}

\affiliation{Department of Physics and Astronomy,
	Uppsala University,\\
	Box 516,
	SE-751 20 Uppsala,
	Sweden}

\emailAdd{alexander.soderberg.rousu@gmail.com}

\makeatletter
\gdef\@fpheader{}
\makeatother

\abstract{
	Replica twist defects are of codimension two and enter in quantum information when finding the R\'enyi entropy. In particular, they generate n replicas of the bulk conformal field theory. We study the monodromy of such defect and learn how a global $O(N)$-symmetry is broken. By applying the equation of motion to the bulk-defect operator-product expansion we are able to extract the anomalous dimension of defect-local fields.

	All of the results in this paper was first presented in my thesis \cite{SoderbergRousu:2023ucv}, and they generalize previous results on an $O(N)$-flavoured monodromy twist defect.
}

\begin{document} 
	
\newtheorem{defin}{Definition}
\newtheorem{thm}{Theorem}
\newtheorem{cor}{Corollary}
\newtheorem{pf}{Proof}
\newtheorem{nt}{Note}
\newtheorem{ex}{Example}
\newtheorem{ans}{Ansatz}
\newtheorem{que}{Question}
\newtheorem{ax}{Axiom}

\maketitle

\section{Introduction}

% Add something regarding generlalized global symmetries?

\textit{Quantum field theories} (QFT's) can in general admit extended objects which covers an entire submanifold $\Sigma^p \subset \M^d$ of the theory (with $p \geq 1$). We call such $p$-dimensional operator a \textit{defect}. Defects will not satisfy the same symmetries as local operators. Properties (like scaling dimensions) of local operators change if they are localized to the defect. This naturally gives rise to an effective theory for adsorption where particles are glued onto a submanifold, see \cite{HURRICKS1970389} and references therein. We thus have to differ between the space outside of the defect, called the \textit{bulk} (or the ambient space), and the space along the defect itself
\begin{equation}
\begin{aligned}
x^\m = x_\para^a \oplus x_\perp^i \ ,
\end{aligned}
\end{equation}
where $\m \in \{0, ..., d - 1\}$, $a \in \{0, ..., p - 1\}$ and $i\in \{1, ..., d - p\}$ (if the time-axis is parallel to the defect).

It is physically important to study defects for various reasons. In condensed matter they give rise to new critical phenomena near the defects. In particular, they can be used to describe impurities in materials (where the underlying microscopic structure of the material may differ). This makes defects important when studying the Kond\={o} effect (scatterings of electrons in metals due to an impurity) \cite{10.1143/PTP.32.37, Affleck:1995ge}.

%Spacetime symmetries (Poincar\'e or conformal) are explicitly broken into a subgroup parallel and orthogonal to it. 
A \textit{homogeneous QFT} (without a defect) has $ISO(d - 1, 1)$-symmetry, and in the conformal case $O(d, 2)$. A defect of dimension $p$ will be invariant under transformations orthogonal to it. In the case of a flat or a spherical defect (the minimal amount of symmetry breaking caused by a defect), this symmetry group will be $SO(d - p)$ if the time-axis is parallel to the defect
\begin{equation}
\begin{aligned}
ISO(d - 1, 1) \rightarrow SO(d - p) \ ,
\end{aligned}
\end{equation}
and $SO(d - p - 1, 1)$ if the time-axis is orthogonal to the defect. In Euclidean signature there is no time direction, and thus the defect enjoy $SO(d - p)$-symmetry.

A \textit{defect-local operator} will satisfy a subgroup of the bulk symmetries, with one part describing transformations along the defect, and the other orthogonal to it (the same as the defect itself). The orthogonal symmetry group will act as a global symmetry group for the defect-local fields. A defect is said to be \textit{conformal} if defect-local operators satisfy conformal symmetry in $p$ dimensions, i.e.  $O(p, 2) \times SO(d - p)$ if the time-axis is parallel to the defect
\begin{equation} \label{DCFT symm break}
\begin{aligned}
O(d, 2) \rightarrow O(p, 2) \times SO(d - p) \ ,
\end{aligned}
\end{equation}
and $O(p + 1, 1) \times SO(d - p - 1, 1)$ if the time-axis is orthogonal to it.\footnote{In two dimensions we can consider one-dimensional defects. Then the group of rotations around the defect is trivial: $SO(d - p) = SO(1) = 1$. In such case only one copy of the extended Virasoro symmetry \cite{PhysRevD.1.2933} is preserved by the defect: $\text{Vir} \times \overline{\text{Vir}} \rightarrow \text{Vir}$.} In Euclidean signature the corresponding symmetry group is $O(p + 1, 1) \times SO(d - p)$. If both the bulk and the defect is conformal, we call the theory a \textit{defect conformal field theory} (DCFT).

Global symmetries are not necessarily broken by the defect. However, they can be, and this can either occur spontaneously or explicitly. A global symmetry might also break due to the monodromy of a codimension two defect ($p = d - 2$). Such defects are special in the sense that they can carry a monodromy action for the fields \cite{Billo:2013jda, Gaiotto:2013nva}, which means that a bulk field transforms under the global symmetry group as we transport it around the defect. This leads to a symmetry breaking of the global symmetry in the bulk. This was studied in \cite{Soderberg:2017oaa} for an $O(N)$-model, where the monodromy action is
\begin{equation} \label{Monodromy constraint}
\begin{aligned}
\ph^i(x_\para, r, \thet + 2\,\pi) = g^{ij}\ph^j(x_\para, r, \thet) \ , \quad g^{ij} \in O(N) \ .
\end{aligned}
\end{equation}
Here we use polar coordinates for the normal directions: $r > 0$, $\thet \in [0, 2\,\pi)$. Due to this constraint, we refer to these defects as \textit{($O(N)$-flavoured) monodromy twist defects} (or symmetry defects). The choice of the group element $g^{ij}$ is referred to as the \textit{twist} and characterizes the defect. In fact, above monodromy relation describes a branch cut in the plane of the normal coordinates. We can choose to treat the monodromy defect as a defect of codimension one which spans a half-plane along this branch cut \cite{Billo:2013jda}.

This monodromy constraint can be  generalized to branch cuts with $n$ Riemann sheets
\begin{equation}
\begin{aligned}
\ph^i_a(x_\para, r, \thet + 2\,\pi) = g^{ij}\ph^j_{a + 1}(x_\para, r, \thet) \ , \quad g^{ij} \in O(N) \ ,
\end{aligned}
\end{equation}
where $\ph_a^i$, with $a\in\{1, ..., n\}$ and $i\in\{1, ..., N\}$, is the fundamental scalar on the $a^{th}$ sheet. The \textit{replica twist defect} (or R\'enyi defect) is the defect with above monodromy where the QFT on each sheet are all the same. They enter in QFT's when applying the replica trick used to find the R\'enyi entropy \cite{Calabrese:2004eu, Casini:2009sr}
\begin{equation}
\begin{aligned}
S_n = \frac{\log Z(n) - n\log Z(1)}{1 - n} \stackrel{n\rightarrow 1}{\longrightarrow} S_{EE} \ ,
\end{aligned}
\end{equation}
where $Z(n)$ is the partition function for the theory with $n$ replicas. In the $n\rightarrow 1$ limit we find the entanglement entropy, $S_{EE}$, in a QFT, which loosely speaking is a measure on how much information of the total system is preserved on a subregion of the full space of the theory \cite{Calabrese:2004eu, Casini:2009sr}.	

If we consider $n$ replicas of the bulk theory, we should identify the $(n + 1)^{th}$ bulk field with the first one. This gives us the monodromy constraint for a general $\ph^i_a$
\begin{equation}
\begin{aligned}
\ph^i_{n + 1} \equiv \ph^i_1 \qRq \ph^i_a(x_\para, r, \thet + 2\,\pi\,n) = g^{ij_1}\prod_{m = 2}^{n}g^{j_{m - 1}j_m}\ph^{j_m}_{a}(x_\para, r, \thet) \ .
\end{aligned}
\end{equation}
Since $O(N)$ is closed, $g^{ij_1}\prod_{m = 2}^{n}g^{j_{m - 1}j_m} \in O(N)$. To avoid clutter we thus write above monodromy constraint as
%\newpage
\begin{equation} \label{replica monodromy}
\begin{aligned}
\ph^i_a(x_\para, r, \thet + 2\,\pi\, n) = g^{ij}\ph^j_{a}(x_\para, r, \thet) \ , \quad g^{ij}\in O(N) \ .
\end{aligned}
\end{equation}
We refer to $g^{ij}$ as the twist of the replica twist defect.

In this paper we study how this monodromy breaks the $O(N)$-symmetry. We do this using the \textit{defect operator-product expansion} (DOE): the OPE between a bulk-local field and the defect itself. For a codimension two defect in a DCFT, it is given by \cite{Gaiotto:2013nva}
\begin{equation} \label{codim 2 DOE}
\begin{aligned}
\ph^i(x) &= \sum_{\hO_s}\mu^{\ph^i}{}_{\hO_s}\frac{e^{i\,s\,\thet}}{r^{\De - \hD_s}}\hat{C}(r^2\p_\para^2)\hO_s(x_\para) \ , \quad \hat{C}(x) &= \sum_{m\geq 0}\frac{x^m}{(-4)^mm! \left( \hD - \frac{d - 4}{2} \right)_m} \ ,
\end{aligned}
\end{equation}
Here $s$ is the $SO(2)$-charge of the defect-local field $\hO_s$. These are all primaries (annihilated by the generators of the special conformal transformations along the defect), and the differential operator, $\hat{C}(r^2\p_\para^2)$, generates the towers of descendants. 

%Note that the exchanged fields does not have any $SO(p - 1, 1)$-spin. Such spinning operators only appear in the DOE of a bulk-local field with non-trivial $SO(d - 1, 1)$-spin, say $l$, wherein such case the DOE contains defect-local operators with $SO(p - 1, 1)$-spin $\hat{l} \leq l$ \cite{Billo:2016cpy, Lauria:2018klo}. 

By applying the DOE to both sides of the monodromy constraint \eqref{replica monodromy}, we find how the global symmetry group is broken along the defect. This was first studied for a $O(1)$($ = \Z_2$)-flavoured monodromy twist defect in \cite{Gaiotto:2013nva}. In the case of $O(N)$ with $N \geq 2$, this was first studied in \cite{Soderberg:2017oaa}, and further studied in \cite{Giombi:2021uae}. In Sec. \ref{Sec. Mono} we will see that this breaking will occur in the same way for \eqref{replica monodromy}. The difference lies in the charges, $s$, of $\hO_s$, which now depend on the number of replicas, $n$.

Furthermore, in Sec. \ref{Sec: eom} we show how the anomalous dimension of the defect-local fields can be extracted using the e.o.m. and the \textit{defect operator product expansion} (DOE). This idea was first developed in \cite{Rychkov:2015naa} where it was applied to the OPE in a homogeneous CFT. It was generalized to a codimension one defect in a free scalar theory in \cite{Gliozzi:2015qsa}, and later to a boundary in an interacting theory in \cite{Giombi:2020rmc} (see also Sec. 3.2 in \cite{Dey:2020jlc}). For a monodromy twist defect \eqref{Monodromy constraint} this method has been used in \cite{Yamaguchi:2016pbj} for a $\Z_2$-twist, and in \cite{Soderberg:2017oaa} for an $O(N)$-twist. This method was generalized in \cite{Nishioka:2022odm, Nishioka:2022qmj} for codimension one and two defects (respectively) to extract the anomalous dimension of defect-local tensorial operators

In Sec. \ref{Sec: eom} we apply it to the $O(N)$-flavoured replica twist defect in a CFT near four dimensions, where we consider a quartic bulk-interaction
\begin{equation} \label{phi 4 thy}
\begin{aligned}
S = \int_{\R^d}d^dx \left( \frac{(\p_\m\ph^i)^2}{2} + \frac{\la}{8}\ph^4 \right) \ .
%\frac{\la^0}{8}\ph^4
\end{aligned}
\end{equation}
Here $\ph^4 \equiv [(\ph^i)^2]^2$ and summation over the group indices, $i \in \{1, .., N\}$, is implicit. The anomalous dimension of $\hO_s$ is found (upto first order in the coupling constant), which gives us a new result which reduces down to that in \cite{Soderberg:2017oaa} when we only consider one replica ($n = 1$).

% In Ch. \ref{Ch: Monodromy} we will study how this monodromy breaks the global $O(N)$-symmetry along the defect. This generalizes the result in \cite{Soderberg:2017oaa}, though it makes use of the same method. In this Chapter we will also extract the anomalous dimensions in $d = 4 - \e$ of the defect-local fields (upto first order in the coupling constant) by applying the e.o.m. to the DOE \eqref{codim 2 DOE}.

%---------------------------------------------------

\section{Monodromy of replica twist defects} \label{Sec. Mono}

We will consider a replica twist defect in an $O(N)$-model, where there are $n$ replicas of the bulk theory. Its monodromy action is given by \eqref{replica monodromy}. By conjugation, a general $O(N)$-element is given by
\begin{equation} \label{twist}
\begin{aligned}
(g^{ij}) = \diag(R_1^\pm, ..., R_p^\pm, \pm 1) \ ,
\end{aligned}
\end{equation}
where the last $O(1) = \Z_2$-element is not present if $N$ is even, and $R_\al^\pm$, $\al\in\{1, ..., p\}$, is a general $O(2)$-matrix characterized by an angle $\vartheta_a$
\begin{equation} \label{R}
\begin{aligned}
R_\al^\pm = \begin{pmatrix}
\pm\cos\vartheta_\al & \mp\sin\vartheta_\al \\
\sin\vartheta_\al & \cos\vartheta_\al
\end{pmatrix} \ , \quad \vartheta_\al \in [0, 2\pi) \ .
\end{aligned}
\end{equation}
Here $R_\al^+ \in SO(2) \subset O(2)$ is a proper rotation (with determinant one), and $R_\al^- \in O(2)$ an improper one (with determinant minus one). We allow each element in the twist \eqref{twist} to differ in this aspect ($\pm$). 

To understand how the global $O(N)$-symmetry is broken along the defect we make use of the DOE \eqref{codim 2 DOE}. The idea is to study which values of $s$ is valid for the monodromy \eqref{replica monodromy} to hold. From this we can find the subgroups of $O(N)$ under which defect-local fields might be charged under. 

Due to the form of the twist \eqref{twist} we need to study the cases $R_1^\pm$ and $\pm 1$ separately. We will start with the latter
\begin{equation}
\begin{aligned}
\ph^N_a(x_\para, r, \thet + 2\,\pi\,n) = \pm \ph^N_a(x_\para, r, \thet) \ .
\end{aligned}
\end{equation}
If we now apply the DOE \eqref{codim 2 DOE} to both sides of this equation we find
\begin{equation} \label{SO(2) charge}
\begin{aligned}
e^{2\,\pi\,i\,n\,s} = \pm 1 \qRq s = \frac{m_\pm}{2\,n} \in \Q \ .
%\sum_{\hO_j}\mu^{\ph}{}_{\hO_j}(s)\frac{e^{-i\,s\,\thet - 2\,\pi\,i\,n\,s}}{r^{\De_\ph - \hD}}\hat{C}_2(r^2\p_\para^2)\hO_j^s(x_\para) = \pm\sum_{\hO_j}\mu^{\ph}{}_{\hO_j}(s)\frac{e^{-i\,s\,\thet}}{r^{\De_\ph - \hD}}\hat{C}_2(r^2\p_\para^2)\hO_j^s(x_\para) \ .
\end{aligned}
\end{equation}
Here $m_\pm$ is even/odd for $\pm 1$ respectively.

Let us now proceed with $R_1^\pm$, where we find the system of equations
\begin{equation}
\begin{aligned}
\left\{ \begin{array}{l l}
\ph^1_a(x_\para, r, \thet + 2\,\pi\,n) = \pm\cos\vartheta_1\ph^1_{a}(x_\para, r, \thet) \mp\sin\vartheta_1\ph^2_{a}(x_\para, r, \thet) \ , \\
\ph^2_a(x_\para, r, \thet + 2\,\pi\,n) = \sin\vartheta_1\ph^1_{a}(x_\para, r, \thet) + \cos\vartheta_1\ph^2_{a}(x_\para, r, \thet) \ .
\end{array} \right.
\end{aligned}
\end{equation}
We now use the DOE \eqref{codim 2 DOE} and compare powers of $r$. This is the same as comparing the terms including the same defect-local operators $\hO_s(x_\para)$ on the two sides.\footnote{Since the anomalous dimensions are included in $\hD_s$ it is safe to assume no mixing.} We find
\begin{equation} \label{monodromy equation}
\begin{aligned}
\left\{ \begin{array}{l l}
%e^{2\,\pi\,i\,n\,s}\mu^{\ph^1}{}_{\hO^{j_1...j_k}_s} = \pm\cos\vartheta_1\mu^{\ph^1}{}_{\hO^{j_1...j_k}_s} \mp\sin\vartheta_1\mu^{\ph^2}{}_{\hO^{j_1...j_k}_s} \ , \\
%e^{2\,\pi\,i\,n\,s}\mu^{\ph^1}{}_{\hO^{j_1...j_k}_s} = \sin\vartheta_1\mu^{\ph^1}{}_{\hO^{j_1...j_k}_s} + \cos\vartheta_1\mu^{\ph^2}{}_{\hO^{j_1...j_k}_s} \ .
e^{2\,\pi\,i\,n\,s}\mu^{\ph^1}{}_{\hO_s} = \pm\cos\vartheta_1\mu^{\ph^1}{}_{\hO_s} \mp\sin\vartheta_1\mu^{\ph^2}{}_{\hO_s} \ , \\
e^{2\,\pi\,i\,n\,s}\mu^{\ph^1}{}_{\hO_s} = \sin\vartheta_1\mu^{\ph^1}{}_{\hO_s} + \cos\vartheta_1\mu^{\ph^2}{}_{\hO_s} \ .
\end{array} \right.
\end{aligned}
\end{equation}
When one of the trigonometric functions in $R_1^\pm$ vanish (i.e. $\vartheta_1 \in \{0, \frac{\pi}{2}, \pi, \frac{3\pi}{2}\}$), we find charges in the same class as \eqref{SO(2) charge}. This is expected as in such case we have \\
$R_1^\pm = \diag(\pm 1, \pm 1)$ ($\pm$ signs not related) after conjugation.

Otherwise, the first equation is generally solved by
\begin{equation} \label{DOE coefficients}
\begin{aligned}
%\mu^{\ph^1}{}_{\hO^{j_1...j_k}_s} = \mp\frac{\sin\vartheta_1}{e^{2\,\pi\,i\,n\,s} \mp \cos\vartheta_1}\mu^{\ph^2}{}_{\hO^{j_1...j_k}_s} \ ,
\mu^{\ph^1}{}_{\hO_s} = \mp\frac{\sin\vartheta_1}{e^{2\,\pi\,i\,n\,s} \mp \cos\vartheta_1}\mu^{\ph^2}{}_{\hO_s} \ .
\end{aligned}
\end{equation}
Then from the second equation of \eqref{monodromy equation} we find
\begin{equation} \label{s eq}
\begin{aligned}
(e^{2\,\pi\,i\,n\,s} - \cos\vartheta_1)(e^{2\,\pi\,i\,n\,s} \mp \cos\vartheta_1) = \mp\sin\vartheta_1^2 \ .
\end{aligned}
\end{equation}
This equation has different solutions for $s$ depending on whether $R_1^\pm$ is proper or improper.  In the proper case we find
\begin{equation}
\begin{aligned}
(e^{2\,\pi\,i\,n\,s} - \cos\vartheta_1)^2 = -\sin\vartheta_1^2 \qRq e^{2\,\pi\,i\,n\,s} = \cos\vartheta_1 \pm i\sin\vartheta_1 \ ,
\end{aligned}
\end{equation}
which has the solution
\begin{equation} \label{SO(2) charge 2}
\begin{aligned}
s = \frac{m}{n} + \frac{\vt_1}{2\,\pi\,n} \ , \quad m\in\Z \ .
\end{aligned}
\end{equation}
If we insert this charge back into \eqref{DOE coefficients} we find\footnote{This relation tells us that the DOE coefficient, $\m^\Ph{}_{\hO_s}$, for the complex scalar $\Ph = \ph^1 + i\,\ph^2$ (transforming under $U(1)$) is real-valued.}
\begin{equation} \label{DOE coeff rel}
\begin{aligned}
%\mu^{\ph^1}{}_{\hO^{j_1...j_k}_s} = \pm i\,\mu^{\ph^2}{}_{\hO^{j_1...j_k}_s} \ .
\mu^{\ph^1}{}_{\hO_s} = \pm i\,\mu^{\ph^2}{}_{\hO_s} \ .
\end{aligned}
\end{equation}
The corresponding two defect-local operators both have $SO(2)$-charge given by \eqref{SO(2) charge 2}.
%Since at least one of these DOE coefficients is complex, it tells us that if a defect-local operator with $SO(2)$-charge \eqref{SO(2) charge 2} appear in the DOE \eqref{codim 2 DOE 2}, then the $(d - 2)$-dimensional CFT on the defect is non-unitary.

Let us now move on to the improper solution of \eqref{s eq}. In such case
\begin{equation}
\begin{aligned}
(e^{2\,\pi\,i\,n\,s} - \cos\vartheta_1)(e^{2\,\pi\,i\,n\,s} + \cos\vartheta_1) = \sin\vartheta_1^2 \qRq e^{4\,\pi\,i\,n\,s} = \pm 1 \ .
\end{aligned}
\end{equation}
This is solved by \eqref{SO(2) charge}. It corresponds to one defect-local operators with even $SO(2)$-charge (w.r.t. $m_\pm$), and another with odd charge.
\iffalse
If we insert this solution back into \eqref{codim 2 DOE 2} we find
\begin{equation}
\begin{aligned}
\mu^{\ph^1}{}_{\hO^{j_1...j_k}_s} = \frac{\sin\vartheta_1}{\pm 1 + \cos\vartheta_1}\mu^{\ph^2}{}_{\hO^{j_1...j_k}_s} \ .
\end{aligned}
\end{equation}
\fi

To summarize:
\begin{itemize}
	\item The $\Z_2$-element in \eqref{twist} gives us one defect operator with either even or odd (w.r.t. $m_\pm$) $SO(2)$-charge \eqref{SO(2) charge}.
	\item Each proper $R_a^+$ gives us a pair of defect operators with the fractional charge \eqref{SO(2) charge 2}. Their DOE coefficients are related by \eqref{DOE coeff rel}.
	\item Lastly, each improper $R_a^-$ gives us one defect operator with even (w.r.t. $m_\pm$) charge \eqref{SO(2) charge}, and another defect operator with odd charge.
\end{itemize}
With this information at hand we can see how the bulk $O(N)$-symmetry is broken along the defect by counting the number of defect fields with the same $SO(2)$-charge. That is, the DOE splits into several different sums, where each sum runs over different classes of $SO(2)$-charges. If we assume that in total there exist: 
\begin{itemize}
	\item $n_+$ defect fields with charge $s \in \frac{\Z}{n}$,
	\item $n_-$ with $s \in \frac{\Z}{n} + \frac{1}{2\,n}$,
	\item $2\,n_1$ with $s \in \frac{\Z}{n} + \frac{\vt_1}{2\,\pi\, n}$,
	\item $2\,n_2$ with $s \in \frac{\Z}{n} + \frac{\vt_2}{2\,\pi\, n}$, \\
	\vdots
	\item $2\,n_q$ with $s \in \frac{\Z}{n} + \frac{\vt_q}{2\,\pi\, n}$,
\end{itemize}
where $0 \leq n_\pm \leq N$, $0 \leq n_r \leq \frac{N}{2}$, $r \in \{1, ..., q\}$ and $0 \leq q \leq p$ with $p$ from the twist \eqref{twist}, then the $O(N)$-symmetry is broken down to
\begin{equation} \label{symm breaking}
\begin{aligned}
O(N) \rightarrow O(n_+) \times O(n_-) \times O(2\,n_1) \times ... \times O(2\,n_q) \ .
\end{aligned}
\end{equation}
% orthogonal
This means that the defect-local fields are in irreducible representations of these subgroups of $O(N)$.
%one (and only one) of these subgroups of $O(N)$.

%\newpage

\section{Anomalous dimensions from the equation of motion} \label{Sec: eom}

Let us now extract the anomalous dimensions of the defect-local operators from the DOE \eqref{codim 2 DOE} using the e.o.m. (in the process we will also find the DOE coefficients in the free theory). In $d = 4 - \e$ we can consider a quartic bulk-interaction \eqref{phi 4 thy}. The bulk coupling has the non-trivial \textit{Wilson-Fisher} (WF) \textit{fixed point} (f.p.) \cite{wilson1973quantum}
\begin{equation} \label{phi 4 fp}
\begin{aligned}
\la^* = \frac{(4\,\pi)^2\e}{N + 8} + \mco(\e^2) \ ,
\end{aligned}
\end{equation}
which gives us the following e.o.m. at the conformal f.p.
\begin{equation} \label{eom}
\begin{aligned}
\p_\m^2\ph^i = \la^*(\ph^j)^2\ph^i \ , \quad \la^* = \frac{8\,\pi^2\e}{N + 8} + \mco(\e^2) \ .
\end{aligned}
\end{equation}
This yields the following \textit{Dyson-Schwinger} (DS) equation
\begin{equation} \label{def DS}
\begin{aligned}
\p_{y^\m}^2\langle\hO_s(x_\para)\ph^i(y)\rangle = \la^*\langle\hO_s(x_\para)(\ph^j)^2\ph^i(y)\rangle \ .
\end{aligned}
\end{equation}
We will start by studying the free theory where RHS of this equation is zero. For simplicity we will consider the bulk-defect correlator, which using the DOE \eqref{codim 2 DOE} can be written as
\begin{equation}
\begin{aligned}
\langle\hO_s(x_\para)\ph^i(y)\rangle &= \m^{\ph^i}{}_{\hO_s}e^{i\,s\,\thet}\sum_{m\geq 0}\frac{a_{\hD_s,m}}{r^{\De_\ph - \hD_s - 2\,m}}\p_\para^{2\,m}\frac{A_d}{|s_\para|^{2\,m}} \ , \\
a_{\hD_s,m} &= \frac{1}{(-4)^mm!\left( \hD_s - \frac{d - 4}{2} \right)_m} \ .
\end{aligned}
\end{equation}
The LHS of the classical DS eq. \eqref{def DS} is then
\begin{align} \label{def LHS}
&\p_\m^2\langle\hO_s(x_\para)\ph^i(y)\rangle = (\p_\para^2 + \p_r^2 + r^{-1}\p_r + r^{-2}\p_\thet^2)\langle\ph^i(x)\hO_s(y_\para)\rangle \nonumber \\
&\quad= \m^{\ph^i}{}_{\hO_s}e^{i\,s\,\thet}\sum_{m\geq 0}\frac{a_{\hD_s, m - 1} + (\hD_s - \De_\ph + 2\,m + s)(\hD_s - \De_\ph + 2\,m - s)a_{\hD_s, m}}{r^{\De_\ph - \hD_s - 2(m - 1)}}\times \nonumber \\
\tq\times \p_\para^{2\,m}\frac{A_d}{|s_\para|^{2\,m}} \ .
\end{align}
In the free theory this is to be zero on its own. By comparing powers of $r$ and avoiding trivial solutions, we find
\begin{equation} \label{eom2}
\begin{aligned}
\frac{(\hD_s - \De_\ph)^2 - 2\,m(2(\De_\ph + 1) - d) - s^2}{(-4)^mm!(\hD_s - \frac{d - 2}{2})_m} = 0 \ , \quad \De_\ph = \frac{d - 2}{2} \ .
\end{aligned}
\end{equation}
This has several different solutions. The Pochhammer symbol in the denominator is zero if
\begin{equation}
\begin{aligned}
\hD_s = \De_\ph - k - m \ , \quad k \in \Z_{\geq 1} \ ,
\end{aligned}
\end{equation}
which is only valid for non-unitary theories on the defect (as it violates the unitary bound in $p = d - 2$ dimensions). Another solution is when the numerator of \eqref{eom2} is zero
\begin{equation} \label{def scaling dim 0}
\begin{aligned}
\hD_s = \De_\ph \pm s \ .
\end{aligned}
\end{equation}
%This result coincide with that of the monodromy twist defect ($n = 1$) \cite{Gaiotto:2013nva, Soderberg:2017oaa}.

\subsection{Dyson-Schwinger equation}

We will now move on to the interacting theory, and study the RHS of the DS eq. \eqref{def DS}
\begin{equation} \label{rhs}
\begin{aligned}
&\la^*\langle\hO_s(x_\para)(\ph^j)^2\ph^i(y)\rangle = (N + 2)\la^*\langle\ph^2(y)\rangle\langle\hO_s(x_\para)\ph^i(y)\rangle + \mco(\e^2) \ .
\end{aligned}
\end{equation}
In order to extract CFT data from this equation, we need the one-point function of $\ph^2$ (at $\mco(\e^0)$), which is found from the coincident-limit of the $\ph - \ph$ correlator. This correlator can be found from the Klein-Gordon equation in radial coordinates
\begin{equation*} %\hspace{-10px}
\begin{aligned}
(\p_\para^2 + \p_{r_1}^2 + r_1^{-1}\p_{r_1} + r_1^{-2}\p_{\thet_1}^2)D^{ij}(s_\para, r_1, r_2, \vph) = r_1^{-1}\de^{ij}\de^{(d - 2)}(s_\para)\de(r_1 - r_2)\de(\vph) \ ,
\end{aligned}
\end{equation*}
where $D^{ij}(s_\para, r_1, r_2, \vph) \equiv \langle\ph^i(x_\para, r_1, \thet_1)\ph^j(y_\para, r_2, \thet_2)\rangle$ and $\vph \equiv \thet_2 - \thet_1$. Details on how this differential equation is solved are in App. B of \cite{Lauria:2020emq}
\begin{equation} \label{corr}
\begin{aligned}
D^{ij}(s_\para, r_1, r_2, \vph) &= A_d\de^{ij}\sum_s\frac{\G_{\hD_s}}{\G_{\De_\ph}\G_{\hD_s - \De_\ph + 1}}\frac{e^{i\,s\,\vph}}{(r_1\, r_2)^{\De_\ph}}(4\,\x)^{-\hD_s}\times \\
\eq\times {}_2F_1\left( \hD_s, \hD_s - \frac{p - 1}{2}, 2\,\hD_s - p + 1, -\x^{-1} \right) \ .
\end{aligned}
\end{equation}
Here $\hD_s$ is given by 
\begin{equation} \label{def scaling dim}
\begin{aligned}
\hD_s = \De_\ph + |s| \ ,
\end{aligned}
\end{equation}
which is a subclass of the free theory solution \eqref{def scaling dim 0}, and the cross-ratio is
\begin{equation}
\begin{aligned}
\x = \frac{s_\para^2 + (r_1 - r_2)^2}{4\,r_1r_2} \ .
\end{aligned}
\end{equation}
Before we find $\langle\ph^2\rangle$, let us extract the DOE coefficients. We do this by comparing \eqref{corr} to the expression found from the DOE \eqref{codim 2 DOE}
\begin{equation}
\begin{aligned}
D^{ij}(s_\para, r, r', \de\thet) &= \sum_s(\mu^{\ph^i}{}_{\hO_s})^\dagger\mu^{\ph^j}{}_{\hO_s}\frac{e^{i\,s\,\vph}}{(r_1r_2)^{\De_\ph}}\frac{A_d}{|s_\para|^{2\,\hD_s}} \ .
\end{aligned}
\end{equation}
If we know expand \eqref{corr} in $r_1, r_2$ we find
\begin{equation}
\begin{aligned}
(\mu^{\ph^i}{}_{\hO_s})^\dagger\mu^{\ph^j}{}_{\hO_s} = \de^{ij}\frac{\G_{\hD_s}}{\G_{\De_\ph}\G_{\hD_s - \De_\ph + 1}} \ .
\end{aligned}
\end{equation}
These are on the same form as in \cite{Gaiotto:2013nva, Soderberg:2017oaa}, with the difference lying in the $SO(2)$-spin $s$.

Now we will proceed with finding $\langle\ph^2\rangle$. In $d = 4$ the summand simplify using the following ${}_2F_1$-identity
\begin{equation} \label{2F1 id}
\begin{aligned}
{}_2F_1\left( |s| + \frac{1}{2}, |s| + 1, 2\,|s| + 1, -\x^{-1} \right) = 4^{|s|}\sqrt{\frac{\x}{\x + 1}}\frac{\x^{|s|}}{(\sqrt{\x} + \sqrt{\x + 1})^{2|s|}} \ .
\end{aligned}
\end{equation}
The coincident-limit, $r_2 \rightarrow r_1 \equiv r$, $\vph \rightarrow 0$, of \eqref{corr} is then
\begin{equation}
\begin{aligned}
D^{ij}(s_\para, r, r, 0) &= \frac{A_d\de^{ij}}{|s_\para|\sqrt{s_\para^2 + 4\,r^2}}\sum_s\left( \frac{2\,r}{|s_\para| + \sqrt{s_\para^2 + 4\,r^2}} \right)^{|s|} \ .
\end{aligned}
\end{equation}
After the change in variables $s = \frac{t}{n}$ (s.t. $t \in \Z + \up$ with $\up \equiv \frac{\vartheta}{2\pi}$), this is resummed in the same way as in App. B.2 of \cite{Soderberg:2017oaa}. We find
\begin{equation}
\begin{aligned}
D^{ij}(s_\para, r, r, 0) &= \frac{2\,A_d\de^{ij}}{|s_\para|\sqrt{s_\para^2 + 4\,r^2}} \frac{\left( \frac{2\,r}{|s_\para| + \sqrt{s_\para^2 + 4\,r^2}} \right)^{\frac{2\,\up}{n}}}{1 - \left( \frac{2\,r}{|s_\para| + \sqrt{s_\para^2 + 4\,r^2}} \right)^{\frac{2}{n}}} \\
&= A_d\de^{ij} \left( \frac{n}{s_\para^2} - \frac{2\,\up - 1}{2\,r\, |s_\para|} + \frac{6\,\up(\up - 1) - n^2 + 1}{12\,n\,r^2} + \mco(s_\para) \right) \ .
\end{aligned}
\end{equation}
We can identify one-point functions of a bulk-local operator $\mco$ (times $|s_\para|^{\De - 2\,\De_\ph}\la^{\ph\ph}{}_{\mco}\m^{\mco}{}_\1$) by comparing above expansion with the bulk OPE. The first term corresponds to the identity exchange, the second to the bulk scalar $\ph$ itself (since the bulk $O(N)$-symmetry is broken by the defect)
\begin{equation}
\begin{aligned}
\langle\ph^i(x_\para, r, \thet)\rangle &= \frac{A_d\m^{\ph}{}_\1\de^{iN}}{r^2} + \mco(\e) \ , \quad \m^{\ph}{}_\1 = \frac{1}{2} - \up \ ,
\end{aligned}
\end{equation}
and the last one to $\ph^2$ which we are interested in%\footnote{Here we used that $\la^{\ph^i\ph^j}{}_{\ph^2} = \de^{ij}$, see e.g. paper \ref{paper: BCFT bootstrap}.}
\begin{equation}
\begin{aligned}
\langle\ph^2(x_\para, r, \thet)\rangle &= \frac{A_d\m^{\ph^2}{}_\1}{r^2} + \mco(\e) \ , \quad \m^{\ph^2}{}_\1 = \frac{6\,\up(\up - 1) - n^2 + 1}{12\,n} \ .
\end{aligned}
\end{equation}
As a sanity check we see that this reduces to the result in \cite{Soderberg:2017oaa} when there is only one replica: $n = 1$.

\subsection{Anomalous dimensions}

We are now ready to find the anomalous dimensions of the defect-local operators, $\hO_s$, that appear in the DOE \eqref{codim 2 DOE} using the DS eq. \eqref{def DS}. 
% $\hO^{j_1...j_k}$, that appear in the DOE \eqref{codim 2 DOE 2} using the e.o.m. \eqref{eom}. 
The RHS of the DS eq. \eqref{rhs} is given by
\begin{equation}
\begin{aligned}
&\langle\hO_s(x_\para)\la(\ph^j)^2\ph^i(y)\rangle = \m^{\ph^i}{}_{\hO_s}e^{i\,s\,\thet}\sum_{m\geq 0}\frac{(N + 2)A_d\m^{\ph^2}{}_\1a_{\hD_s,m}\la}{r^{\De_\ph - \hD_s - 2(m - 1)}}\p_\para^{2\,m}\frac{A_d}{|s_\para|^{2\,m}}  \ .
%\langle\hO^{j_1...j_k}(x_\para)\la(\ph^j)^2\ph^i(y)\rangle = \m^{\ph^i}{}_{\hO^{j_1...j_k}}e^{is\thet}\sum_{m\geq 0}\frac{(N + 2)A_d\m^{\ph^2}{}_\1a_{\hD,m}\la}{r^{\De_\ph - \hD - 2(m - 1)}}\p_\para^{2m}\frac{A_d}{|s_\para|^{2m}}  \ .
\end{aligned}
\end{equation}
Powers in $r$ can now be compared to the LHS of \eqref{def DS}, i.e. \eqref{def LHS}, which gives us
\begin{equation*}
\begin{aligned}
a_{\hD_s, m - 1} + (\hD_s - \De_\ph + 2\,m + s)(\hD_s - \De_\ph + 2\,m - s)a_{\hD_s, m} = (N + 2)A_d\m^{\ph^2}{}_\1a_{\hD_s,m}\la \ .
\end{aligned}
\end{equation*}
Let us now expand both sides in $\e$ (where we use as input that the bulk anomalous dimension, $\g_\ph$, start first at $\mco(\e^2)$ \cite{wilson1973quantum})
\begin{equation} \label{defect scaling dim}
\begin{aligned}
\De_\ph = \frac{d - 2}{2} + \mco(\e^2) \ , \quad \hD_s = \frac{d - 2}{2} + |s| + \e\,\hga_s + \mco(\e^2) \ ,
\end{aligned}
\end{equation}
which yields the anomalous dimension of the defect-local fields
\begin{equation} \label{defect anom dim}
\begin{aligned}
\hga_s = \frac{N + 2}{N + 8}\frac{\m^{\ph^2}{}_\1}{|s|} = \frac{N + 2}{N + 8}\frac{6\,\up(\up - 1) - n^2 + 1}{12\,n\,|s|} \ .
\end{aligned}
\end{equation}
This is a new result, which reduces down to the result in \cite{Soderberg:2017oaa} when $n = 1$.

\section{Conclusion}

In this work we showed that the replica twist defect with monodromy action \eqref{replica monodromy} breaks a global $O(N)$-symmetry in the same way \eqref{symm breaking} as the monodromy defect in \cite{Soderberg:2017oaa}. We found that the difference lies in the defect-local operators, $\hO_s$, which now has $SO(2)$-charge (\ref{SO(2) charge}, \ref{SO(2) charge 2}) (depending on which subgroup of $O(N)$ it transforms under). At the conformal f.p. \eqref{eom} near four dimensions \eqref{phi 4 thy} we found the scaling dimension of $\hO_s$ to be given by (\ref{defect scaling dim}, \ref{defect anom dim}) upto $\mco(\e)$.

There are several interesting directions to further pursue studies. It would be interesting to find the anomalous dimension of $\hO_s$ for different models. E.g. near three dimensions with a sextic interaction \cite{duplantier1982lagrangian}, or near six dimensions with cubic interactions \cite{10.1143/PTP.54.1828, Fei:2014yja}. To use the method in Sec. \ref{Sec: eom} we need $\langle\ph^2\rangle$. However, if we are not near four dimensions we cannot use the ${}_2F_1$-identity \eqref{2F1 id}, making the calculations more difficult.
% the one-point function of $\ph^2$

We could also follow the lines of \cite{Nishioka:2022qmj}, and apply the techniques from Sec. \ref{Sec: eom} to extract the anomalous dimension of tensorial operators localized to the defect. A possible challenge with this approach is the existence of \textit{multiple-copy operators} in the bulk. These are composite operators of fields from different replicas, e.g. $\ph_a\ph_{a + 1}$ and $\ph_a\ph_{a + 1}\ph_{a + 2}$.

Alternatively we can study the DOE of the bulk Noether current corresponding to $O(N)$, and try find the tilt operator on the defect corresponding to the \textit{conformal manifold} $O(N)/O(n_+) \times O(n_-) \times O(2\,n_1) \times ... \times O(2\,n_q)$ (the broken part of the symmetry group) \cite{bray1977critical, Cuomo:2021cnb, Padayasi:2021sik}. See also Sec. 2.2.1 in my thesis \cite{SoderbergRousu:2023ucv}. In turn, from the four-point function of the protected tilt operator we can read off the curvature of the conformal manifold \cite{Drukker:2022pxk, Herzog:2023dop}.

This tilt operator is also of great importance in numerical bootstrap \cite{Gimenez-Grau:2022czc}, wherein an $O(N)$-flavoured monodromy twist defect \eqref{Monodromy constraint} was considered. This defect has also been analytically conformally bootstrapped in \cite{Gimenez-Grau:2021wiv}, where Lorentzian inversion formulas for the bulk- and defect-channels were used. It would be interesting to apply similar bootstrap techniques (either numerically or analytically) to the replica twist defect \eqref{replica monodromy}.

\section*{Acknowledgement}

I would like to express my gratitude to Agnese Bissi, Pietro Longhi and Andrea Manenti for enriching discussions on the replica twist defects. Moreover, I thank everyone that went to my public defence of my thesis \cite{SoderbergRousu:2023ucv}, where the results in this paper was first presented. This project was funded by Knut and Alice Wallenberg Foundation grant KAW 2021.0170, VR grant 2018-04438 and Olle Engkvists Stiftelse grant 2180108.

\iffalse

\appendix

\section{Appendix}

\fi

\bibliographystyle{utphys}
\footnotesize
\bibliography{References}	
	
\end{document}